\documentclass[aps,prd,twocolumn,groupedaddress,showpacs,preprintnumbers]
{revtex4-1}

\usepackage[dvips]{graphicx}
\usepackage{bm}
\usepackage{amsmath,amssymb}
\allowdisplaybreaks

\newcommand{\cl}{\text{cl}}
\newcommand{\fm}{\text{fm}}

\newcommand{\p}{\partial}
\newcommand{\tr}{\mathop{{\rm tr}}}
\newcommand{\Tr}{\mathop{{\rm Tr}}}
\newcommand{\calL}{\mathcal{L}}
\newcommand{\calI}{\mathcal{I}}
\newcommand{\calJ}{\mathcal{J}}

\newcommand{\drv}[2]{\frac{d #1}{d #2}}
\newcommand{\EQ}{\text{EQ}}
\newcommand{\MeV}{\text{MeV}}
\newcommand{\VEV}[1]{\langle #1\rangle}

\newcommand{\half}{\frac12}
\newcommand{\abs}[1]{\left| #1\right|}

\begin{document}

\preprint{KUNS-2254}

\affiliation{Department of Physics, Kyoto University, Kyoto 606-8502,
Japan}

\title{Relativistic Collective Coordinate Quantization of Solitons:
Spinning Skyrmion}

\author{Hiroyuki Hata}
\email[]{hata@gauge.scphys.kyoto-u.ac.jp}
\author{Toru Kikuchi}
\email[]{kikuchi@gauge.scphys.kyoto-u.ac.jp}
\affiliation{
Department of Physics, Kyoto University, Kyoto 606-8502, Japan}


\begin{abstract}
We develop a consistent relativistic generalization of
collective coordinate quantization of field theory solitons.
Our principle of introducing collective coordinates is that the
equations of motion of the collective coordinates ensure those of the
original field theory.
We illustrate this principle with the quantization of spinning degrees
of freedom of Skyrmion representing baryons.
We calculate the leading relativistic corrections to the static
properties of nucleons, and find that the corrections are
non-negligible ones of $10\%$ to $20\%$.
\end{abstract}


\maketitle

\section{Introduction}
Collective coordinates of a field theory soliton describe the motion
of the soliton in the symmetry directions of the field theory action.
They correspond to the zero modes around the soliton and are
important in understanding its low energy dynamics.
The simplest way of introducing the collective coordinates is to
promote the (originally constant) parameters of a static
soliton to time-dependent dynamical variables by discarding the
fluctuation of nonzero modes.
This is valid in the nonrelativistic cases where time derivatives of
the collective coordinates are small enough.

However, in certain circumstances, it is inappropriate to assume
that solitons move slowly; relativistic corrections come to
be important.
A typical example is the spinning Skyrmion representing baryons
\cite{SkyrmeI}.
In the standard quantization of the rotational collective
coordinate of the Skyrmion \cite{ANW,AN}, the mass of the baryons
is given as the sum of the energy of the static classical solution and
the rotational kinetic energy of a spherical rigid body.
For the nucleon (delta), about $8\%$ ($30\%$) of the total mass comes
from the rotational energy.
This leads to the picture of a baryon rotating with a velocity
comparable to the light velocity at the radius of order $1\,\fm$.
Since the baryon rotates so fast, it is natural to expect that the
baryon cannot maintain its original spherical shape.
The collective coordinate of rotation should be
introduced in a consistent way that can express the deformation.

The purpose of this paper is, on the basis of a general and simple
principle of introducing the collective coordinates,
to carry out the quantization of rotational degrees of freedom of
the $SU(2)$ Skyrmion to evaluate the relativistic corrections to the
static properties of nucleons.
While several authors have discussed the deformation of spinning
Skyrmions with various physical pictures (see Refs.\
\cite{Bander:1984gr}--\cite{Li:1987kq} for earlier works), we
emphasize that the deformation is naturally induced in our treatment
of collective coordinates.
Another feature of our treatment of rotational collective
coordinate is that we introduce it through coordinate-transformed
static solitons with new coordinate depending on the
time derivative of the collective coordinate
[see Eq.\ \eqref{eq:U=Ucl(bmy)} together with \eqref{eq:OSS}].

\section{General principle}
Our basic idea is to avoid mismatch between the field theory dynamics
and the collective coordinate dynamics.  While the system of
collective coordinates has a finite number of degrees of freedom, the
original field theory has an infinite one.  Therefore, even when the
equation of motion (EOM) of collective coordinates holds, it does not
necessarily mean that the soliton configuration satisfies its field
theory EOM. Then our general principle of introduction of collective
coordinates is simply stated as follows:
{\em Collective coordinates must be introduced in such a way that the
  EOM of the collective coordinates ensures that of the original field
  theory.}
If the original field theory is a relativistic one, this principle
should automatically lead to a relativistic dynamics of collective
coordinates.

One way to realize this principle is, starting with suitably
introduced collective coordinates, to integrate over the nonzero
modes around the soliton. Namely, we solve the EOM of the nonzero
modes to express them in terms of the collective coordinates.
Since the field theory EOM is equivalent to the set of EOMs of both
the zero and the nonzero modes, solving the EOM for the latter should
lead to a system of collective coordinates satisfying our principle.
In fact, the relativistic energy of the center-of-mass motion,
$E=\sqrt{P^2+M^2}$, is obtained by this method in scalar field
theories in $1+1$ dimensions \cite{GJS,GJS2}.
However, no explicit calculation by this method has been carried out
for more complicated cases other than the center-of-mass, in
particular, for the collective coordinate of rotational motion.

In this paper, we propose another way: First, the collective
coordinates are put into the static classical solution in such a way
as to fulfill our principle of EOMs. Then, the (relativistic)
Lagrangian of the collective coordinates is obtained simply by
inserting this soliton field into the field theory Lagrangian density
and carrying out the space integration. In this process, nonzero
modes do not appear; in other words, the EOM of nonzero modes implies
that they are equal to zero in this framework.

In the case of the collective coordinate $\bm{X}(t)$ of the
center-of-mass of a soliton in a scalar field theory, we can show that
our principle is satisfied, up to $O(\p_0^6)$ terms, by the
relativistic replacement of the space coordinate $\bm{x}$ of the
static soliton $\varphi^\cl(\bm{x})$ by
$\left(\bm{x}-\bm{X}(t)\right)/\sqrt{1-\bm{V}^2}$ with
$\bm{V}=\dot{\bm{X}}$, leading to the relativistic Lagrangian
$L=-M\sqrt{1-\bm{V}^2}$.
On the other hand, it is a nontrivial task to realize our principle
for the collective coordinate of space rotation.
In the rest of this paper, we illustrate our general method with
two-flavor spinning Skyrmion to obtain the relativistic corrections to
the earlier results \cite{ANW,AN}.
Our finding here is that our principle of EOMs can be satisfied, up to
$O(\p_0^4)$ terms, by taking the coordinate-transformed static soliton
given by \eqref{eq:U=Ucl(bmy)} and \eqref{eq:OSS} with suitably chosen
functions $A(r)$, $B(r)$, and $C(r)$.

Although we focus on the spinning collective coordinate of the
Skyrmion in the rest of this paper, we wish to emphasize that our
method of extracting the system of collective coordinates on the basis
of the principle of EOMs is simpler and has wider applicability since
we are not bothered by the nonzero modes.
In addition, our method is interesting also in that we can directly
know how the soliton deforms due to fast collective motion, since the
collective coordinates are introduced by deforming the coordinate of
the static solution.

\section{The Skyrmion}
The $SU(2)$ Skyrme model \cite{SkyrmeI} is
described by the chiral Lagrangian with the Skyrme term:
\begin{equation}
\calL=\tr\left\{
-\frac{f_\pi^2}{16}L_\mu^2 +\frac{1}{32e^2}[L_\mu,L_\nu]^2
+ \frac{f_\pi^2}{8} m_\pi^2 \left(U-\bm{1}_2\right)\right\},
\label{eq:calL}
\end{equation}
where $U(x)$ is an $SU(2)$ matrix and $L_\mu=-iU\p_\mu U^\dagger$.
The EOM reads
\begin{equation}
\p_\mu\left(L^\mu -\frac{1}{e^2f_\pi^2}[L_\nu,[L^\mu,L^\nu]]\right)
\!-im_\pi^2\left(U-\half\tr U\right)=0.
\label{eq:FTEOM}
\end{equation}
This theory has a static soliton solution
$U^\cl(\bm{x})=\exp\bigl(i \bm{x}\cdot\bm{\tau} F(r)/r\bigr)$
called Skyrmion, where $\tau_i$ are sigma matrices and $F$ is a
function of $r=|\bm{x}|$.
This static solution has rotational collective coordinate, that is,
$U^{\cl}(R^{-1}\bm x)$ is also an solution for any time-independent
$SO(3)$ matrix $R$. For the Skyrmion, the space rotation is equivalent
to the isospin rotation;
$U^{\cl}(R^{-1}\bm x)=W U^\cl(\bm{x}) W^{-1}$ with $W$ being the
$SU(2)$ matrix corresponding to $R$.

The standard way \cite{ANW,AN} to quantize these spinning degrees of
freedom is as follows:
Promoting the constant $SO(3)$ matrix $R$ [or equivalently, the
$SU(2)$ matrix $W$] to a time-dependent one, we take
\begin{equation}
U(\bm x,t)=U^\cl\bigl(R^{-1}(t)\bm{x}\bigr),
\label{eq:U=Ucl(Rinvbmx)}
\end{equation}
as the spinning Skyrmion field and insert it into the Lagrangian
density \eqref{eq:calL}.
Carrying out the spatial integration, we get the Lagrangian of $R(t)$,
$L(R,\dot{R})=\int\!d^3x\,\calL\bigl(
U(\bm x,t)=U^\cl(R^{-1}(t)\bm x)\bigr)=-M^\cl+\half\calI\bm{\Omega}^2$,
where the rest mass $M^\cl$ and the moment of inertia $\calI$ are
functionals of $F(r)$, and $\bm{\Omega}$ is the angular velocity
$\Omega_i=\half\varepsilon_{ijk}(R^{-1}\dot{R})_{jk}
=\tr(i\dot{W}W^{-1}\tau_i)$  (precisely speaking, $\bm{\Omega}$ is the
angular velocity in isospace).
Note that $\bm{\Omega}^2=-\half\Tr(R^{-1}\dot{R})^2$.
We carry out the quantization of the dynamical variable $R$ using this
Lagrangian.

This quantization procedure is evidently a nonrelativistic one
since the Lagrangian of $R$ is simply that of a rigid body.
In addition, the field theory EOM \eqref{eq:FTEOM} with
$U$ given by \eqref{eq:U=Ucl(Rinvbmx)} is violated by the
$O(\bm{\Omega}^2)$ term even if we use the EOM of $R$,
\begin{equation}
(d/dt)\bm{\Omega}=0 .
\label{eq:EOM_R}
\end{equation}
Moreover, the relativistic corrections to the various properties of
baryons seem to be important as we explained in the Introduction.

\section{Deformation of spinning Skyrmion}
We wish to give a relativistic version of the spinning Skyrmion field
\eqref{eq:U=Ucl(Rinvbmx)} on the basis of our general principle of
introducing the collective coordinates. However, it is difficult to
find the complete one in a closed form. Instead, in this paper we
present the leading correction to the rigid body approximation
\eqref{eq:U=Ucl(Rinvbmx)} with respect to the power of the angular
velocity $\bm{\Omega}$, or equivalently, the number of
time derivatives. Our spinning Skyrmion takes the form
\begin{equation}
U(\bm{x},t) = U^\cl(\bm{y}),
\label{eq:U=Ucl(bmy)}
\end{equation}
with $\bm{y}$ given by
\begin{align}
\bm{y}&= \left(1+A(r)(\dot R R^{-1}\bm{x})^2
+ r^2 B(r)\Tr(R^{-1}\dot R)^2 \right)R^{-1}\bm{x}
\notag\\
&\quad
+ r^2 C(r) \bigl[(R^{-1}\dot R)^2 R^{-1}\bm{x}\bigr],
\label{eq:OSS}
\end{align}
where the functions $A$, $B$ and $C$ are to be determined to fulfill
our principle concerning the EOMs.
The form of $\bm{y}$ is the most general one which is at most
quadratic in $\bm{\Omega}$, odd under $\bm{x}\to -\bm{x}$, and has the
property that the left (right) constant $SO(3)$ transformation on $R$
induces the space (isospin) rotation.
The last property is the basic one for the quantization of Skyrmion.
It can easily be seen that the field configuration
\eqref{eq:U=Ucl(bmy)} with $\bm{y}$ of \eqref{eq:OSS} represents a
spheroidal one.

\section{Determination of $\bm{(A,B,C)}$}
For determining $A$, $B$ and $C$ from
our principle, we substitute $U(\bm{x},t)$ of \eqref{eq:U=Ucl(bmy)}
into the field theory EOM \eqref{eq:FTEOM} to find that its
left-hand-side is, upon using the EOM of $U^\cl$, given by
\begin{align}
&\left[\left(\drv{}{t}R^{-1}\dot R\right)\bm{y}\right]_i
\left(L_i^\cl-\frac{1}{e^2f_\pi^2}\bigl[L_j^\cl,[L_i^\cl,L_j^\cl]\bigr]\right)
\notag \\
&+r^2\Tr(R^{-1}\dot R)^2(\bm{y}\cdot\bm{\tau})\times\EQ_1
\!+\!(R^{-1}\dot R\bm y)^2(\bm{y}\cdot\bm{\tau})\times \EQ_2
\notag \\
&+r^2\bigl[(R^{-1}\dot R)^2\bm{y}\bigr]\cdot\bm{\tau}\times \EQ_3
\notag \\
&+r\bigl[\bm{y}\times(R^{-1}\dot R)^2\bm{y}\bigr]
\cdot\bm{\tau}\times \EQ_4
+O(\p_0^4),
\label{eq:FTEOM_Ucl(bmy)}
\end{align}
with $L^\cl_i=-iU^\cl(\bm{y})(\p/\p y_i)U^{\cl}(\bm{y})^\dagger$.
$\EQ_i$ ($i=1,\ldots,4$) are linear in $(A,B,C)$ and their
first and second derivatives with respect to $r$ with coefficients
given in terms of $F$ and its derivatives.
Concrete expressions of $\EQ_i$ are very lengthy, and they are given
in the Appendix.
Here, we present a special linear combination of $\EQ_i$ which
consists of only a special combination $Y=-A+3B+C$ and its
derivatives:
\begin{align}
&\EQ_Y=-3\,\EQ_1+\EQ_2-\EQ_3
\notag \\
&=\left(1+\frac{8}{e^2f_\pi^2}\frac{\sin^2F}{r^2}\right)
F'\drv{^2Y}{r^2}
+\biggl\{2F''+\frac{8}{r}F'
\notag \\
&+\frac{8}{e^2f_\pi^2}
\biggl(2\frac{\sin^2F}{r^2}F''
+\frac{\sin2F}{r^2}(F')^2+\frac{6}{r}\frac{\sin^2F}{r^2}F'
\biggr) \biggr\}\drv{Y}{r}
\notag \\
&+\biggl\{\frac{6}{r}F''+\frac{14}{r^2}F'-\frac{2}{r^3}\sin2F
+\frac{8}{e^2f_\pi^2}\biggl(\frac{8}{r}\frac{\sin^2F}{r^2}F''
\notag \\
&+\frac{4}{r^3}\sin2F(F')^2+\frac{6}{r^2}\frac{\sin^2F}{r^2}F'
-\frac{2}{r^3}\frac{\sin^2F}{r^2}\sin2F \biggr)\biggr\}Y
\notag \\
&-\frac{1}{2r^3}\sin2F + \frac{2}{e^2f_\pi^2}\biggl(
\frac{2}{r}\frac{\sin^2F}{r^2}F''
+\frac{1}{r^3}\sin2F(F')^2
\notag \\
&+\frac{4}{r^2}\frac{\sin^2F}{r^2}F'
-\frac{2}{r^3}\sin2F\frac{\sin^2F}{r^2} \biggr),
\label{eq:EQY}
\end{align}
where the prime on $F$ denotes an $r$-derivative.
The function $Y(r)$ is related to the angle average of $\bm y^2$
with respect to $\bm{x}$,
$1/(4\pi)\int\!d\Omega_{\bm x}\,\bm{y}^2
=r^2\left(1-\frac43 r^2\bm{\Omega}^2\,Y(r)\right)$,
and it seems to represent an independent degree of freedom of the
deformation due to spinning.

Returning to \eqref{eq:FTEOM_Ucl(bmy)}, our principle of introducing
the collective coordinates demands that \eqref{eq:FTEOM_Ucl(bmy)}
vanish identically upon using the EOM of $R$.
As we will see later, the EOM of $R$ remains unchanged from
\eqref{eq:EOM_R} even if the relativistic corrections are introduced.
This implies that the {\em three} functions $(A,B,C)$ must satisfy
{\em four} differential equations $\EQ_i=0$ ($i=1,\ldots,4$), which is
apparently overdetermined.
Fortunately, $\EQ_3$ and $\EQ_4$ are not independent;
we have $\EQ_4=-\tan F\,\EQ_3$ [see Eqs.\ \eqref{eq:EQ3} and
\eqref{eq:EQ4}].
Therefore, $(A,B,C)$ are determined by three inhomogeneous linear
differential equations of second order, $\EQ_i=0$ ($i=1,2,3$), or
another independent set
$\EQ_2=\EQ_3=\EQ_Y=0$ with $\EQ_Y$ given by \eqref{eq:EQY}.
The boundary conditions for $(A,B,C)$ at $r=0$ and $r=\infty$ are
chosen to be the least singular ones among those allowed by the
differential equations;
$(A,B,C)\sim(1,1/r^2,1/r^2)$ as $r\to 0$ and
$(A,B,C-A)\sim(1/r,1/r^3,1/r^2)$ as $r\to\infty$, both up to
numerical coefficients.
By this choice, relativistic corrections to
various physical quantities of baryons, which are given as space
integrations with integrands consisting of $(A,B,C)$ and $F$ and their
derivatives, are unambiguously determined.

\section{Lagrangian of $\bm{R}$}
Inserting the spinning field $U(\bm x,t)$ of \eqref{eq:U=Ucl(bmy)}
into the Lagrangian density \eqref{eq:calL} and spatially integrating
it, we obtain the Lagrangian of $R$ of the following form:
\begin{equation}
L=-M^\cl+\half\calI\bm{\Omega}^2+\frac14\calJ\bm{\Omega}^4,
\label{eq:L_R}
\end{equation}
where $\calI$ and $M^\cl$ are the same as in the rigid body
approximation \cite{AN}, and the last $\bm{\Omega}^4$ term represents
our relativistic correction.
This Lagrangian should be regarded as the rotational motion
counterpart of the relativistic Lagrangian of center-of-mass;
$-M\sqrt{1-\bm{V}^2}=-M +\half M\bm{V}^2+\frac18 M\bm{V}^4+\cdots$.
The EOM of $R$ derived from \eqref{eq:L_R} is
\begin{equation}
\drv{}{t}\Bigl[\left(\calI +\calJ\bm{\Omega}^2\right)
\bm{\Omega}\Bigr]=0 .
\end{equation}
This implies \eqref{eq:EOM_R}, which we already used in deriving the
differential equations for $(A,B,C)$.

The coefficient $\calJ$ of the relativistic correction term has two
origins, $\calJ=\calJ_1+\calJ_2$;
$\calJ_1$ is from the part of the Skyrme model Lagrangian
\eqref{eq:calL} quadratic in $L_0$ and hence is linear in $(A,B,C)$,
while $\calJ_2$ is from the part of \eqref{eq:calL} without $L_0$ and
is quadratic in $(A,B,C)$.
The calculation of $\calJ_2$ is complicated, but fortunately we have
the relation $\calJ_1+2\calJ_2=0$. This is understood by making the
replacement $(A,B,C)\to\lambda(A,B,C)$ in \eqref{eq:OSS} and using the
fact that $\lambda=1$ must give an extremal of $L$ for any constant
$\bm{\Omega}$ satisfying the EOM \eqref{eq:EOM_R} due to our principle
of introducing the collective coordinates and hence of determining
$(A,B,C)$. Therefore, we have $\calJ=\calJ_1/2$, which is explicitly
given by
\begin{align}
&\calJ=\frac{4\pi f_\pi^2}{15}\int_0^\infty\!dr\,
r^4\sin^2 F\biggl\{rZ'+5Z-C +\frac{4}{e^2f_\pi^2}
\notag \\
&\times\biggl[
\frac{\sin^2 F}{r^2}\left(rZ'+3Z+2C\right)
-(F')^2\left(rZ'+Z+C\right)\biggr]\bigg\},
\label{eq:calJ}
\end{align}
with $Z=-2A+5B+2C$.

Several comments are in order:
First, the moment of inertia $\calI$ in the Lagrangian \eqref{eq:L_R}
receives no correction from the deformation of \eqref{eq:OSS} as we
mentioned before. A possible correction to $\calI$ is from the part of
\eqref{eq:calL} without $L_0$ and is linear in $(A,B,C)$.
The vanishing of this correction is shown by the same
$\lambda$-rescaling argument as we used in deriving
$\calJ_1+2\calJ_2=0$.
By the same reason, further correction to $\bm{y}$ \eqref{eq:OSS} of
$O(\p_0^4)$ does not affect $\calJ$.

Our second comment is on another way of obtaining the expression
\eqref{eq:calJ} for $\calJ$.
The isospin charge derived from \eqref{eq:L_R} is
\begin{equation}
I_a=\left(\calI+\calJ\bm{\Omega}^2\right)\Omega_a .
\label{I_a}
\end{equation}
On the other hand, $I_a$ is also given by
$I_a=\int\!d^3x\,J_{V,a}^{\mu=0}$,
where $J_{V,a}^{\mu}$ is the Noether current of $SU(2)_V$ symmetry
derived from the Lagrangian density \eqref{eq:calL}.
By comparing the two expressions of $I_a$, we can directly
obtain \eqref{eq:calJ} from the Noether current.
We can show in general that conserved charges derived from the
Lagrangian \eqref{eq:L_R} of the $R$ system and the
corresponding ones derived as the space integration of the
time component of the Noether currents in the Skyrme model agree with
each other up to the EOM of $R$ and that in the Skyrme model.

Finally, if the pion mass $m_\pi$ in \eqref{eq:calL} is zero, the
integration \eqref{eq:calJ} for $\calJ$ diverges at $r=\infty$.
This is the case also for relativistic corrections to other physical
quantities. It is crucial for our analysis to introduce nonzero pion
mass.

\section{Static properties of nucleons}
The Hamiltonian of $R$ obtained
from the Lagrangian \eqref{eq:L_R} is given by
\begin{equation}
H=M^\cl+\half\calI\bm{\Omega}^2+\frac{3}{4}\calJ\bm{\Omega}^4.
\label{eq:H_R}
\end{equation}
For obtaining the value of $H$ for an eigenstate of the isospin, we
have to solve
\begin{equation}
\bm{I}^2=\left(\calI+\calJ\bm{\Omega}^2\right)^2\,\bm{\Omega}^2,
\label{eq:I^2}
\end{equation}
to express $\bm{\Omega}^2$ in terms of a given $\bm{I}^2$.

Following \cite{AN}, we evaluated $f_\pi$ and $e$ in the Skyrme
Lagrangian \eqref{eq:calL} by taking the masses of nucleon, $\Delta$
and pion ($m_N=939\,\MeV$, $m_\Delta=1232\,\MeV$, $m_\pi=138\,\MeV$)
as inputs.
Our result is $f_\pi=125\,\MeV$ and $e=5.64$.
Compared with the experimental value $f_\pi=186\,\MeV$,
our $f_\pi$ is fairly improved from that of \cite{AN},
$f_\pi=108\,\MeV$, without relativistic correction.
We have also obtained the expressions of various static properties of
nucleons with relativistic correction and computed their numerical
values.
For example, the isoscalar mean square charge radius is given by
$
\VEV{r^2}_{I=0}=4\pi\int_0^\infty \!dr\,r^4J_B^{{\rm cl}\,0}(r)
\left(1+\frac{4}{3}\bm{\Omega}^2r^2Y(r)\right),
$
with $J_B^{\cl\,0}(r)=-1/(2\pi^2)(\sin F/r)^2(dF/dr)$ being
the baryon number density of the static Skyrmion $U^\cl(\bm{x})$.
Our numerical results are summarized in Table \ref{tbl:staticP}.

\begin{table}[t]
\caption[]{The static properties of nucleons.
Prediction of this paper and that of Ref.\ \cite{AN} both use
the experimental values of $(m_N,m_\Delta,m_\pi)$ as inputs.
We follow the notations of Ref.\ \cite{ANW}.}
\begin{ruledtabular}
\begin{tabular}{c|cccc}
&\parbox{17mm}{Prediction\\ (this paper)}
&\parbox{15mm}{Prediction\\ (Ref.\ \cite{AN})}
&Experiment
\\ \hline
$f_\pi$ & $125\,\MeV$ & $108\,\MeV$ & $186\,\MeV$
\\ \hline
$\VEV{r^2}_{I=0}^{1/2}$& $0.59\,\fm$ & $0.68\,\fm$ & $0.81\,\fm$
\\
$\VEV{r^2}_{I=1}^{1/2}$& $1.17\,\fm$ & $1.04\,\fm$ & $0.94\,\fm$
\\
$\VEV{r^2}_{M,I=0}^{1/2}$& $0.85\,\fm$ &$0.95\,\fm$ & $0.82\,\fm$
\\
$\VEV{r^2}_{M,I=1}^{1/2}$& $1.17\,\fm$ & $1.04\,\fm$& $0.86\,\fm$
\\
$\mu_p$ & $1.65$ & $1.97$ & $2.79$
\\
$\mu_n$ & $-0.99$ & $-1.24$ & $-1.91$
\\
$\abs{\mu_p/\mu_n}$ & $1.67$ & $1.59$ & $1.46$
\\
$g_A$ & $0.58$ & $0.65$ & $1.24$
\end{tabular}
\end{ruledtabular}
\label{tbl:staticP}
\end{table}

As seen from the table, our relativistic correction is a
non-negligible one of roughly $10\%$ to $20\%$ for every static
property. (Note that each of our numerical values is not given simply
by adding the $\bm{\Omega}^2$ correction to the value of Ref.\
\cite{AN}, since the parameters $f_\pi$ and $e$ themselves are also
changed.)
Unfortunately, the correction is not in the direction of making the
theoretical prediction closer to the experimental value for most
of the quantities. However, we emphasize that this is not a problem of
our basic principle of collective coordinate quantization; it might be
due to taking only the first term of the expansion in powers of
$\bm{\Omega}^2$, or to the fact that the Skyrme model is merely an
approximation to QCD. In relation to the first possibility, the
ratio of the contributions of the three terms of the Hamiltonian
\eqref{eq:H_R} to the baryon mass is approximately $89:7:4$
for the nucleon, while it is $68:14:18$ for $\Delta$. This shows that
our analysis using $\Delta$ needs better treatment of the relativistic
correction beyond simple expansion in powers of angular velocity.

\section{Summary}
In this paper, we proposed a general
principle of introducing collective coordinates of solitons and
applied it to the quantization of spinning motion of
Skyrmion. We computed the leading relativistic corrections
to the Lagrangian of the rotational motion and various physical
quantities of baryons. Compared with the rigid body approximation,
the value of the decay constant $f_\pi$ has become fairly close to
the experimental one due to the correction, but the numerical results
are not good for other static properties of nucleons.
Putting aside the problem of comparison with the experimental values,
our result shows the importance of relativistic treatment of the
spinning collective coordinate beyond the rigid body approximation.

Finally, application of our principle of collective coordinate
quantization to other interesting physical systems is of course
an important future subject.

\begin{acknowledgments}
We would like to thank Kenji Fukushima, Koji Hashimoto, Antal Jevicki
and Keisuke Ohashi for valuable discussions.
The work of H.~H.\ was supported in part by a Grant-in-Aid for
Scientific Research (C) No.\ 21540264 from the Japan Society for the
Promotion of Science (JSPS).  The work of T.~K.\ was supported by
a Grand-in-Aid for JSPS Fellows No.\ 21-951.
The numerical calculations were carried out on Altix3700 BX2 at YITP
in Kyoto University.
\end{acknowledgments}

\appendix

\begin{widetext}
\section{Explicit expressions of $\EQ_{1,2,3,4}$}
\label{sec:EQs}

In this appendix, we present concrete expressions of the four
quantities $\EQ_i$ ($i=1,2,3,4$) appearing in
\eqref{eq:FTEOM_Ucl(bmy)}.
\begin{align}
\EQ_1&=\frac{2}{r^2}F'A-F'\drv{^2B}{r^2}
-2\left(F''+\frac{4}{r}F'\right)\drv{B}{r}
-\frac{2}{r}\left(3F''+\frac{7}{r}F'-\frac{1}{r^2}\sin2F\right)B
-\frac{1}{r^2}\left(2F'-\frac{1}{r}\sin2F\right)C
\notag\\
&
+\frac{1}{f_\pi^2e^2}\Biggl\{\frac{4}{r^4}(1-\cos2F)F'A
-\frac{4}{r^2}(1-\cos2F)F'\frac{d^2B}{dr^2}
-\frac{8}{r^2}\left[(1-\cos2F)\left(F''+\frac{3}{r}F'\right)
+\sin2F\,(F')^2\right]\drv{B}{r}
\notag \\
&-\frac{8}{r^3}\left[
(1-\cos 2F)\left(4F''+\frac{3}{r}F'-\frac{1}{r^2}\sin 2F\right)
+4\sin 2F\,(F')^2\right]B
\notag\\
&-\frac{2}{r^3}(1-\cos2F)F'\drv{C}{r}
-\frac{4}{r^3}\left[
(1-\cos 2F)\left(F''+\frac{2}{r}F'-\frac{1}{r^2}\sin 2F\right)
+\sin 2F\,(F')^2\right]C
\Biggr\} ,
\label{eq:EQ1}
\\
\EQ_2&= -F'\drv{^2A}{r^2}
-2\left(F''+\frac{4}{r}F'\right)\drv{A}{r}
+\frac{2}{r}\left[-3 F''+\frac{1}{r}(\cos2F-3)F'+\frac{1}{r^2}\sin2F
\right]A
\notag \\
&+\left(F'-\frac{1}{2r}\sin2F\right)\drv{^2C}{r^2}
+\left[2F''+\frac{1}{r}(7-\cos2F)F'-\frac{3}{r^2}\sin2F
\right]\drv{C}{r}
+\frac{2}{r}\left[3 F''+\frac{2}{r}(1-\cos2F)F'\right]C
\notag\\
&+\frac{1}{f_\pi^2 e^2}\Biggl\{
\frac{2}{r^3}\left[(1-\cos 2F)\left(
F''+\frac{2}{r}F'-\frac{2}{r^2}\sin 2F\right)
+2\sin 2F\,(F')^2\right]
\notag\\
&-\frac{4}{r^2}(1-\cos2F)F'\drv{^2A}{r^2}
-\frac{4}{r^2}\left[2(1-\cos 2F)\left(F''+\frac{3}{r}F'\right)
+\sin 2F\,(F')^2\right]\drv{A}{r}
\notag \\
&-\frac{4}{r^3}\left[(1-\cos 2F)\left(8F''+\frac{1}{r}(1-2\cos 2F)F'
-\frac{2}{r^2}\sin 2F\right) +5\sin 2F\,(F')^2\right]A
\notag \\
&+\frac{1}{r^2}(1-\cos 2F)\left(4F'-\frac{1}{r}\sin 2F
\right)\drv{^2C}{r^2}
\notag\\
&+\frac{2}{r^2}\left[(1-\cos 2F)\left(4F''+\frac{1}{r}(7-2\cos 2F)F'
-\frac{2}{r^2}\sin 2F\right)+2\sin 2F\,(F')^2\right]\drv{C}{r}
\notag\\
&+\frac{4}{r^3}\left[(1-\cos 2F)\left(5F''-\frac{4}{r}(1+\cos 2F)F'
+\frac{2}{r^2}\sin 2F\right)+4\sin 2F\,(F')^2\right]C\Biggr\} ,
\label{eq:EQ2}
\\
\EQ_3&=2\cos F\times\EQ_{34} ,
\label{eq:EQ3}
\\
\EQ_4&=-2\sin F\times\EQ_{34} ,
\label{eq:EQ4}
\end{align}
where $\EQ_{34}$ in \eqref{eq:EQ3} and \eqref{eq:EQ4} is given by
\begin{align}
\EQ_{34}&=\frac{1}{2r^3}\sin F +\frac{2}{r^2}\cos F\,F'A
-\frac{1}{2r}\sin F\drv{^2C}{r^2}
-\frac{1}{r}\!\left(\cos F\,F'+\frac{3}{r}\sin F\right)\!\drv{C}{r}
-\frac{1}{r^2}\left(4\cos F\,F'+\frac{1}{r}\sin F\right)C
\notag \\
&+\frac{1}{f_\pi^2 e^2}\sin F\Biggl\{
\frac{2}{r^3}\left[(F')^2-\frac{1}{r^2}(1-\cos 2F)\right]
+\frac{4}{r^2}(F')^2\drv{A}{r}
+\frac{4}{r^3}F'\left(3F'+\frac{2}{r}\sin 2F\right)A
-\frac{1}{r^3}(1-\cos2F)\drv{^2C}{r^2}
\notag \\
&-\frac{4}{r^2}\left[(F')^2+\frac{1}{r}\sin 2F\,F'
+\frac{1}{r^2}(1-\cos 2F)\right]\drv{C}{r}
-\frac{4}{r^3}\left[
(F')^2+\frac{4}{r}\sin 2F\,F'-\frac{1}{r^2}(1-\cos 2F)
\right]C\Biggr\} .
\label{eq:EQ34}
\end{align}

\end{widetext}


\end{document}